\documentclass[aps,showpacs,a4paper,floatfix,twocolumn,prc]{revtex4-1}
\usepackage{graphicx,color,xcolor}
\usepackage{amsfonts,amsmath}
\usepackage{amssymb,ulem}
\usepackage{dcolumn}
\usepackage{bm,lipsum}
\usepackage[utf8]{inputenc}
\usepackage{subeqnarray}
\usepackage{array}
\usepackage{epsfig,hyperref}
\usepackage{morefloats}
\usepackage{relsize}
\usepackage{url}

\newcommand{\bwt}{\begin{widetext}}
\newcommand{\ewt}{\end{widetext}}
\newcommand{\beq}{\begin{equation}}
\newcommand{\eeq}{\end{equation}}
\newcommand{\bea}{\begin{eqnarray}}
\newcommand{\eea}{\end{eqnarray}}

\begin{document}

\title{Delta baryons in neutron stars}

\author{Kauan D. Marquez}
\affiliation{Departamento de F\'{\i}sica - CFM, Universidade Federal de Santa Catarina,  Florian\'opolis, SC CEP 88.040-900, CP. 476, Brazil}
\author{Helena Pais}
\affiliation{CFisUC, Department of Physics, University of Coimbra, 3004-516 Coimbra, Portugal}
\author{Débora P. Menezes}
\affiliation{Departamento de F\'{\i}sica - CFM, Universidade Federal de Santa Catarina,  Florian\'opolis, SC CEP 88.040-900, CP. 476, Brazil}
\author{Constan\c{c}a Provid\^encia}
\affiliation{CFisUC, Department of Physics, University of Coimbra, 3004-516 Coimbra, Portugal}

\begin{abstract}
By applying a relativistic mean-field description of neutron star matter with density dependent couplings, we analyse the properties of two different matter compositions: nucleonic matter with $\Delta$ baryons and nucleonic matter with hyperons and $\Delta$ baryons. The delta-meson couplings are allowed to vary within a wide range of values obtained by experimental data, while the  hyperon-meson couplings are fitted to hypernuclear properties. Neutron star properties with no deconfinement phase transition  are studied. It is verified that many models are excluded because the effective nucleon mass becomes zero before the maximum mass configuration is attained. Hyperon-free with $\Delta$-dominated composition compact stars are possible, the \textit{deltic stars}. It is found that with a convenient choice of parameters the existence of deltic stars with 80\% of $\Delta$ baryons at the center of the star is possible. However, the presence of hyperons lowers the $\Delta$ baryon fraction to values below 20\% at the center and below 30\% at 2-3 saturation densities. 
It is discussed that in the presence of $\Delta$ baryons, the hyperon softening is not so drastic because $\Delta$s couple more strongly to the $\omega$ meson, and the stiffness of the equation of state is determined by the $\omega$-dominance at high densities. The speed of sound reflects very well this behavior.  
The compactness of the pulsar {RX J0720.4-3125} 
imposes $x_{\sigma\Delta}>x_{\omega\Delta}>1$ and favors
$x_{\rho\Delta}>1$. 
\end{abstract}

\maketitle

\section{Introduction}
 
 Although the class of stellar remnants that are neither white dwarves nor black holes is traditionally named \textit{neutron stars} (NS), these objects are not composed solely of neutrons. 
 Even the more na\"ive description of such objects must include some amount of protons 
 in order to guarantee the stability of the nuclear matter, and this fact was already pointed out in the first proposals of the existence of NS by Landau, Baade and Zwicky in the early 1930s.
 Almost forty years ago, Glendenning \cite{Glendenning:1984jr} discussed in his seminal paper different scenarios considering non-nucleonic degrees of freedom in NS matter, including hyperons, $\Delta$ baryons, pions and kaons, within a relativistic mean field approach. 
 In this work, Glendenning found that the $\Delta$ baryons do not nucleate inside the NS core. This result was due to the coupling parameters chosen, as it was shown later that, with a convenient choice of the couplings minimally constrained by the existing experimental measurements,  $\Delta$ baryons may indeed occur inside neutron stars
 \cite{Xiang:2003qz,Schurhoff:2010ph,Lavagno:2010ah,Drago:2013fsa,Drago:2014oja,Cai:2015hya,Kolomeitsev:2016ptu,Li:2018qaw,Ribes:2019kno,Raduta:2021xiz}. 
 
 The knowledge of the NS composition and the signatures of this composition is presently a field of intense investigation.
 To consider the entire spin-1/2 baryon octet as part of the NS matter composition is almost the standard in the nuclear astrophysics community
 \cite{Glendenning:1982nc,Prakash:1995uw,Baldo:1999rq,Oertel:2014qza,Vidana:2015rsa,marquez17,Roark:2018boj,Stone:2019blq,Sedrakian:2020kbi,menezes2021neutron,Motta:2022nlj} 
 but, more recently, there is a strong interest in understanding how the presence of the $\Delta$ baryons specifically may influence the properties of NS and their evolution
 \cite{Li:2019tjx,Malfatti:2020onm,Li:2020ias,Thapa:2020ohp,backes2021effects,Thapa:2021kfo,Dexheimer:2021sxs,Raduta:2022elz,Marczenko:2022hyt}.
 The lightest spin-3/2 baryons are just $\sim30\%$ heavier than the nucleons, and are even lighter than the heaviest spin-1/2 baryons of the octet, what makes very reasonable to expect them to appear at the same density range as the hyperons (about 2-3 times the nuclear saturation density). 
One thing that could forbid the $\Delta$ onset would be if they were subject to a very repulsive coupling, but that is not the case, since their coupling potential for isospin-symmetric matter at saturation density is expected to be attractive and in a range of to $2$/$3$ to $1$ times the potential of the nucleons  \citep{Drago:2014oja,Kolomeitsev:2016ptu,Raduta:2021xiz}.

 In \cite{Custodio:2022fbb}, the authors have studied the effect of heavy baryons on the constitution of hot non-homogeneous matter, in particular their effects on the light clusters abundance and dissolution, using two relativistic mean-field nuclear models (FSU2H \cite{Tolos:2017lgv}, a model with non-linear mesonic terms,  and DD2 \cite{Typel2009}, a model with density-dependent couplings). For the $\Delta$ baryon, the couplings were restricted to values compatible with experimental observations as discussed in \cite{Drago:2014oja,Ribes:2019kno}. 
 It was found that the model FSU2H was much more restrictive, because most of the couplings would not be acceptable to describe neutron stars since the effective nucleon mass would become zero at densities below the maximum mass configuration. 
 On the other hand, the DD2 model seemed to show much more flexibility and allowed a wider range of acceptable couplings. 
 In \cite{Ribes:2019kno}, the FSU2H model has been fully investigated, but there was no reference to the implications of the fact that the effective nucleon mass may become null at still low densities.  In \cite{Kolomeitsev:2016ptu}, this problem was also encountered, but the authors have modified their model in order to avoid this issue. 
 
 In the present work, we will explore in depth the effects of the $\Delta$ baryon couplings considering a model that describes adequately nuclear matter properties and NS observations, considering the $\Delta$ admixture, in both pure nucleonic and hyperonic NS matters. 
 We will study the behavior of the nucleon effective mass, that was not addressed in Ref.  \cite{Li:2019tjx}, the speed of sound, the $\Delta$ and hyperonic fraction and the electron chemical potential, and also discuss the star properties such as mass and radius.
We will pay special attention to some interesting aspects, as the possible increase of the NS maximum mass as compared to hyperonic only stars, or the possibility of the formation of stars with more than 80\% of $\Delta$ baryons at the core center. Also, special compact stars may exist in some hyperon-free $\Delta$-dominated composition,  
 referred as {\it deltic stars}.

\section{Formalism}

{In this study, hadronic matter is described  within a relativistic mean-field approach with density-dependent couplings. 
This class of models is shown to be very consistent in the description of nuclear matter experimental properties \cite{dutra2014}, and also when astrophysical constraints are imposed \cite{PhysRevC.93.025806,lourencco2019}.
In such models, the interaction is described through the exchange of mesons, and here we consider the scalar meson $\sigma$, the vector mesons $\omega$ and $\phi$ (that carries hidden strangeness), isoscalars,  and the isovector-vector meson $\vec\rho$.}

{In this approach the Lagrangian density reads as
\begin{align} 
 \label{lagtw}
{\cal L}={}&\sum_b \bar{\Psi}_b  \left[\gamma_\mu\left(i\partial^{\mu}-\Gamma_{\omega b} \omega^{\mu}-\Gamma_{\phi b} \phi^{\mu}-
\frac{\Gamma_{\rho b}}{2}  \vec{\tau} \cdot \vec{\rho}^\mu \right) \right. \nonumber \\ & \left. 
-\left(m_b-\Gamma_{\sigma b} \sigma\right)\right] \Psi_b
+\frac{1}{2}(\partial_{\mu}\sigma\partial^{\mu}\sigma
-m_\sigma^2 \sigma^2) \nonumber \\
&-\frac{1}{4}\Omega_{\mu\nu}\Omega^{\mu\nu}
+\frac{1}{2}m_\omega^2 \omega_{\mu}\omega^{\mu} 
-\frac{1}{4}\Phi_{\mu\nu}\Phi^{\mu\nu}
+\frac{1}{2}m_\phi^2 \phi_{\mu}\phi^{\mu} \nonumber \\
&-\frac{1}{4}\vec R_{\mu\nu}\cdot\vec R^{\mu\nu}+\frac{1}{2}
m_\rho^2 \vec \rho_{\mu}\cdot \vec \rho^\mu \, ,
\end{align}
where $m_i$ is the mass associated with the $i=\sigma,\omega,\phi,\rho$ meson field,
$\Omega_{\mu\nu}=\partial_{\mu}\omega_{\nu}-\partial_{\nu}\omega_{\mu}$,
$\Phi_{\mu\nu}=\partial_{\mu}\phi_{\nu}-\partial_{\nu}\phi_{\mu}$
$\vec R_{\mu\nu}=\partial_{\mu}\vec \rho_{\nu}-\partial_{\nu} \vec\rho_{\mu}
- \Gamma_\rho (\vec \rho_\mu \times \vec \rho_\nu)$, and
$\vec{\tau}$ is the isospin matrix (with vectors in isospin space denoted by arrows). The sum on the index $b$ runs over all the baryonic species considered in the matter composition, described by the field $\psi_b$ with the mass $m_b$. We are aware that spin-3/2 baryons (as the $\Delta$s) are in fact described by the Rarita-Schwinger Lagrangian density, which would demand a discrimination on \eqref{lagtw} for the terms when $b=\{\Delta\}$.
 Nonetheless, the resulting equation of motion can be written compactly as a Dirac equation in the mean-field approximation,  see \cite{dePaoli:2012eq}.

The density-dependent coupling constants $\Gamma_{\sigma}$,
$\Gamma_{\omega}$ and $\Gamma_{\rho}$ are adjusted in order to reproduce some of the nuclear matter bulk properties using the following scaling with the baryonic density $n_B$
\begin{equation}
\Gamma _{i}(n_B)=\Gamma _{i}(n_0)a_{i}\frac{1+b_{i}(\eta+d_{i})^{2}}{1+c_{i}(\eta+d_{i})^{2}}
\label{paratw1}
\end{equation}
for $i=\sigma,\omega$ and
\begin{equation}
\Gamma _{\rho}(n_B)=\Gamma _{\rho}\exp [-a_{\rho }(\eta-1)],  \label{paratw2}
\end{equation}
with $\eta=n_B/n_0$, where $n_0$ is the nuclear saturation density.
The Euler-Lagrange equations are used to calculate the equations of motion for the meson and baryon fields, see for instance \cite{Typel1999}, and a complete description for the hadronic matter given by this Lagrangian density can be derived from there. 
The model parameters considered here are obtained from a fitting that considered known experimental constraints on values of nuclear matter binding energy, compressibility modulus, symmetry energy and its slope,  as well the $^{208}$Pb neutron skin measurements. This parameterization is labeled as DDME2 and its details can be found in \cite{Lalazissis2005}.} {In Table \ref{tab1}, we give its symmetric nuclear matter properties at saturation density.}
\begin{table}[htb]
\caption{\label{tab1}
{The symmetric nuclear matter properties at saturation density for the DDME2 model: the nuclear saturation density $n_0$, the binding energy per particle $B/A$, the incompressibility $K$, the symmetry energy $E_{sym}$, the slope of the symmetry energy $L$, and the nucleon effective mass $M^{*}$. All quantities are in MeV, except for $n_0$ that is given in fm$^{-3}$, and the effective nucleon mass is normalized to the nucleon mass.}}
\begin{ruledtabular}
\vspace{0.5cm}
\begin{tabular}{ccccccc}
 Model    &  $n_0$ & $B/A$ &  $K$ & $ E_{sym}$ & $L$ & $M^{*}/M$\\
\hline
DDME2   & 0.152   & 16.14   & 251  &  32.3 & 51 & 0.57 \\
\end{tabular}
\end{ruledtabular}
\end{table}

The fitting of the model free parameters are made by considering ordinary nuclear matter, composed only by nucleons (protons and neutrons). 
We parametrize the other baryonic species couplings in terms of the nucleon-meson couplings by defining the ratio $x_{ib}=\Gamma_{ib}/ \Gamma_{i}$, with $i =\sigma,\omega,\phi,\rho$ and $b=\{N\},\{H\},\{\Delta\}$, where it is defined that $x_{iN}=1$.
The hyperon couplings are defined taking the same density-dependence of the couplings to the $\sigma$ and $\omega$ mesons as the one of the nucleons, and for the density-dependence of the $\phi$ coupling we take the one of the $\omega$ meson.  
The couplings of the $\sigma$ meson to the $\Lambda$ and $\Xi$ hyperons were determined from a fit to  hypernuclear  {binding energies, and were taken from \cite{Fortin:2017cvt}  for the $\Lambda$,  and from \cite{Fortin:2020qin} for the $\Xi$} 
, while for the $\Sigma$ hyperon, it has been fixed by imposing that at saturation the $\Sigma$ potential in symmetric nuclear matter is $+30$ MeV, i.e. we have considered a repulsive interaction as seems to be the indication from experimental measurements \cite{Gal:2016boi}.
The couplings to the $\sigma$ meson have been taken from \cite{Fortin:2017cvt,Fortin:2020qin}
$$x_{\sigma\Lambda}=0.621,\quad x_{\sigma\Sigma}=0.467,\quad  x_{\sigma\Xi}=0.321.$$
The magnitude of the couplings  for the isoscalar-vector mesons are given by the SU(6) symmetry
\begin{align*}
    &x_{\omega\Lambda}=x_{\omega\Sigma}=\frac{2}{3},\quad
x_{\omega\Xi}=\frac{1}{3},\\
&x_{\phi\Lambda}=x_{\phi\Sigma}=-\frac{\sqrt{2}}{3},\quad
x_{\phi\Xi}=-\frac{2\sqrt{2}}{3}.
\end{align*}
The coupling  of each hyperon to the $\rho$ meson is defined by the product of the hyperon isospin with the $\rho$ meson coupling to the nucleon, i.e., 
$x_{\rho H}=\tau_H$.

The $\Delta$-meson couplings are treated rather freely in this study.
Some experimental constraints summarized in \cite{Drago:2014oja} suggest that the nucleon-$\Delta$ potential is slightly more attractive than the nucleon-nucleon one, which consequently implies $x_{\sigma\Delta}\geq1$. Also, the vector coupling is constrained by results \cite{WEHRBERGER1989797} as respecting the relation
\begin{equation}
    0\leq x_{\sigma\Delta}-x_{\omega\Delta}\leq 0.2, \label{eq:constr}
\end{equation}
with no constraint put in the $x_{\rho\Delta}$ value.
All of these constraints will be taken with a grain of salt, as we aim to explore the behavior of NS matter according to this parameters in a comprehensive way, not discarding the whole regions of possible values beforehand. 
 These constraints will be remembered in the evaluation of the results. 
Early investigations on the behavior of these parameters were made in \cite{de2000delta,deltaslo2,Ribes:2019kno}, but no previous study analyzed carefully the astrophysical implications of the {vanishing}
nucleon effective mass, 
among other considerations.

In order to describe NS matter properly, we must observe charge neutrality and chemical equilibrium conditions. To reach these constraints, a non-interacting gas of leptons (electrons and muons) is included in the description. At zero temperature, the particle fractions can be determined from the neutron and electron chemical potentials, such that the particle fractions $y_i=n_i/n_B$ are determined from the neutron and electron chemical potentials through
\begin{equation}
    \mu_b=\mu_n-q_b\mu_e, \label{betaeq}
\end{equation}
where $q_b$ is the electric charge of the baryon $b$, and $\mu_\mu=\mu_e$.
{ In the inner crust of the star, nonspherical clusters may form, the so-called nuclear pasta phases. For this EoS region, we consider a self-consistent Thomas-Fermi calculation \cite{grill14} with the same RMF model and under the same thermodynamic conditions as for the homogeneous gas core, i.e., $\beta-$equilibrium matter at zero temperature. This inner crust EoS has recently been published \cite{DDME2composeInner} in the CompOSE database \cite{compose}. For the outer crust, we use the EoS by Baym, Pethick and Sutherland \cite{BPS}, that was added below baryon density of 0.0003 fm$^{-3}$. The unified inner-crust--core DDME2 EoS can also be found in CompOSE \cite{DDME2composeUnif}.
}

\section{Results and discussions}

\begin{figure}[!t]
\centering
\includegraphics[width=.95\linewidth]{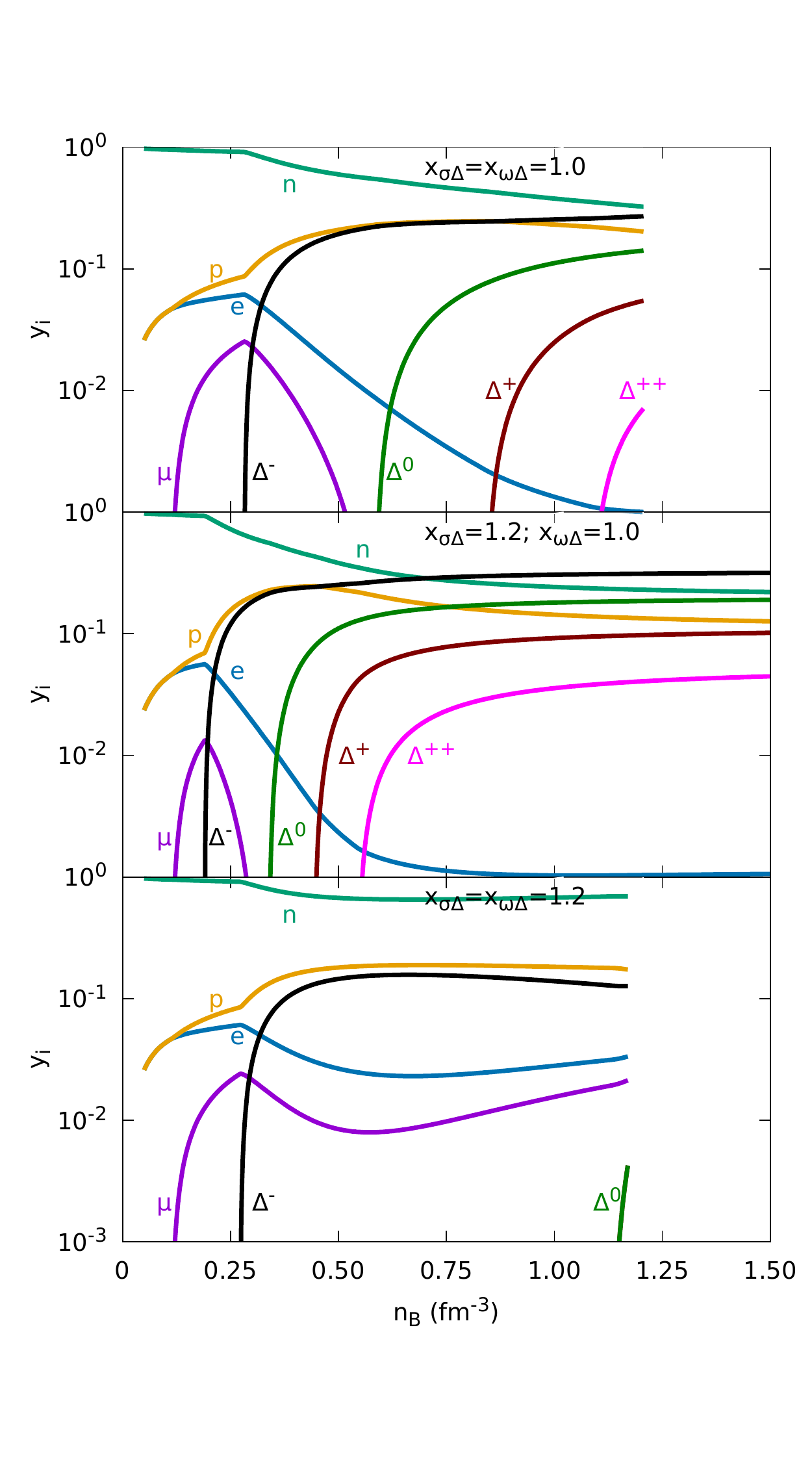} 
\caption{Particle relative populations as a function of the density, fixing $x_{\rho\Delta}=1.0$, for the N$\Delta$ matter composition. {The particle fraction curves stop at the vanishing effective mass density (see text).}
}\label{fig:population_nd}
\end{figure}

\begin{figure}[!t]
\centering
\includegraphics[width=.95\linewidth]{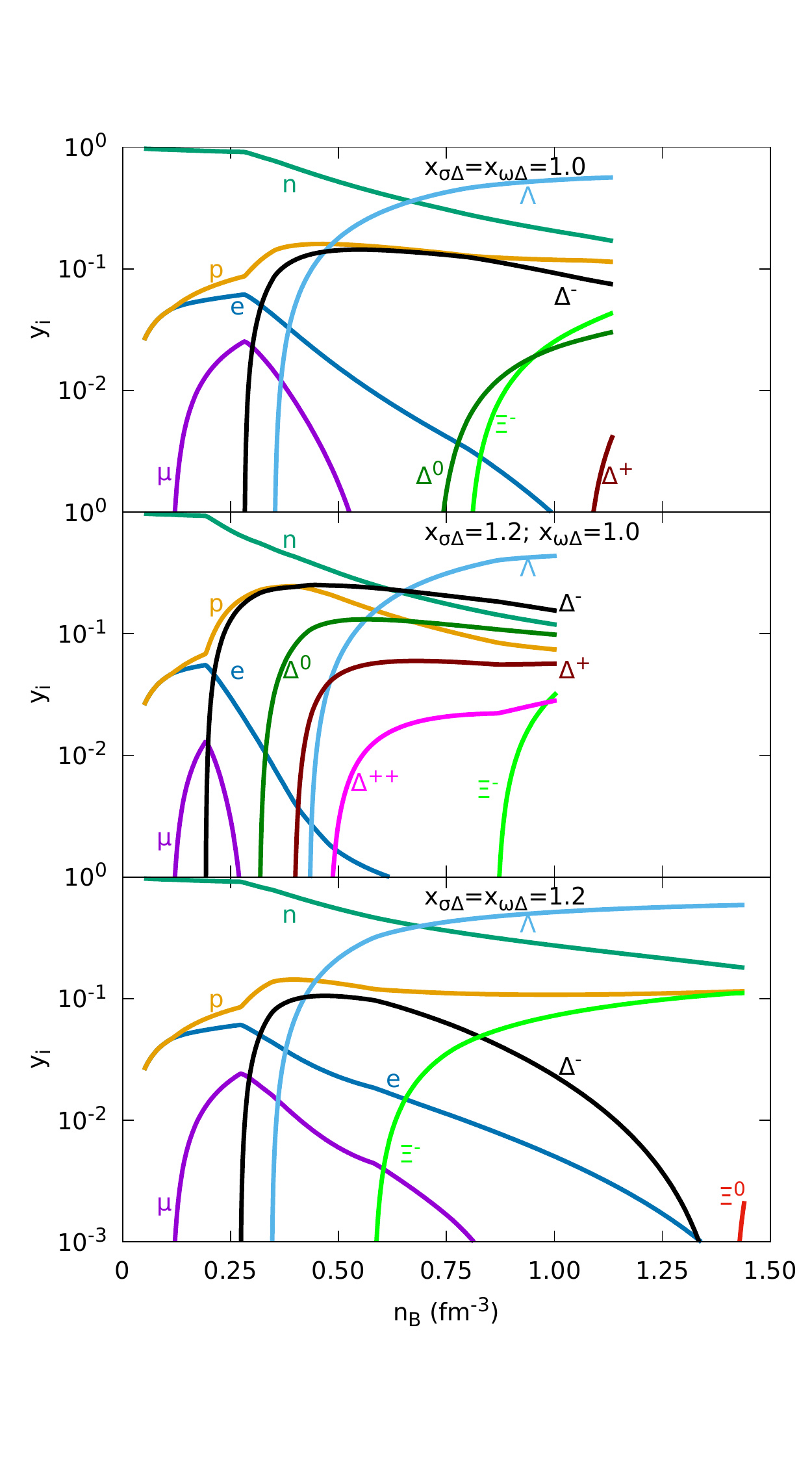}
\caption{Same as Fig. \ref{fig:population_nd}, but for the  NH$\Delta$ matter composition. 
}\label{fig:population_nhd}
\end{figure}

We first discuss the composition and expected onset of the different heavy baryons in $\beta$-stable, charge-neutral NS matter, as described by the DDME2 model formalism exposed in the previous section. 
Figs. \ref{fig:population_nd} and \ref{fig:population_nhd} show the particle fractions when the baryonic composition considered is the hyperon-free matter, i.e., composed by nucleons and $\Delta$ baryons (labeled N$\Delta$), and $\Delta$-admixed hypernuclear matter, i.e., composed by nucleons, hyperons and $\Delta$s (labeled NH$\Delta$). 
The negatively charged spin-$3/2$ baryons are favored when charge neutrality is enforced, while the positively charged ones are suppressed, in the same way as what usually takes place with the hyperons.
Being negatively charged, the $\Delta^{-}$ can replace a neutron-electron pair at the top of their Fermi seas, being favored over the lighter $\Lambda$ and $\Sigma$ baryons because of the fact that their potential is more attractive, to a proportion which the mass difference is counterbalanced. When allowed, the first hyperon to appear is the $\Lambda$, as it is the lighter one and neutrally charged.

{Analysing Figs. \ref{fig:population_nd} and \ref{fig:population_nhd},  we  conclude  that if the coupling fractions are lager than one, in hyperon free (N$\Delta$) matter, having $x_{\sigma\Delta}>x_{\omega\Delta}$ favors the appearance of all $\Delta$ species, even the one with charge +2;
if $x_{\sigma\Delta}=x_{\omega\Delta}$, the larger the coupling the less favored the $\Delta$ baryons are, due to the $\omega$-dominance at large densities that occurs because the $\sigma$ field saturates;
also, the electrons are  efficiently replaced by $\Delta^-$ baryons if the $x_{\omega\Delta}$ is not too large.
When hyperons are included in the (NH$\Delta$) matter, the $\Lambda$ hyperon sets after the $\Delta^-$ and is pushed to high densities if $x_{\sigma\Delta}>x_{\omega\Delta}$; if $x_{\sigma\Delta}=x_{\omega\Delta}$ the larger $x_{\omega\Delta}$ the more important is the contribution of the $\Xi^-$ hyperon, and the smaller $\Delta^-$ because the larger the $x_{\omega\Delta}$, the more repulsive the $\Delta^-$ interaction at high densities, see \cite{Ribes:2019kno}. The presence of hyperons strongly disfavors the increase of $\Delta$ fractions at high densities because hyperons feel a much weaker repulsion since the coupling to the $\omega$ meson is smaller.
This fact is exemplified by the competition between the $\Delta^-$ and the $\Xi^-$, as one can notice in the bottom panel of Fig. \ref{fig:population_nhd}, whith the former suppressing the first as it is lighter and subject to a less repulsive coupling.}
For a fixed $x_{\omega\Delta}$, there will always be a $x_{\sigma\Delta}$ where the $\Delta^{-}$ and the $\Lambda$ appear at the same density, beyond which the resonances are favored.
{Moreover, as already discussed in \cite{Ribes:2019kno,Drago:2014oja}, if hyperons are explicitly included and  the constraint given by Eq. \eqref{eq:constr} is imposed, $\Delta^{-}$ will always appear first, shifting the onset of hyperons to larger densities compared with the $\Delta$-free threshold density.}

\begin{figure*}[!t]
\centering
\includegraphics[width=\linewidth]{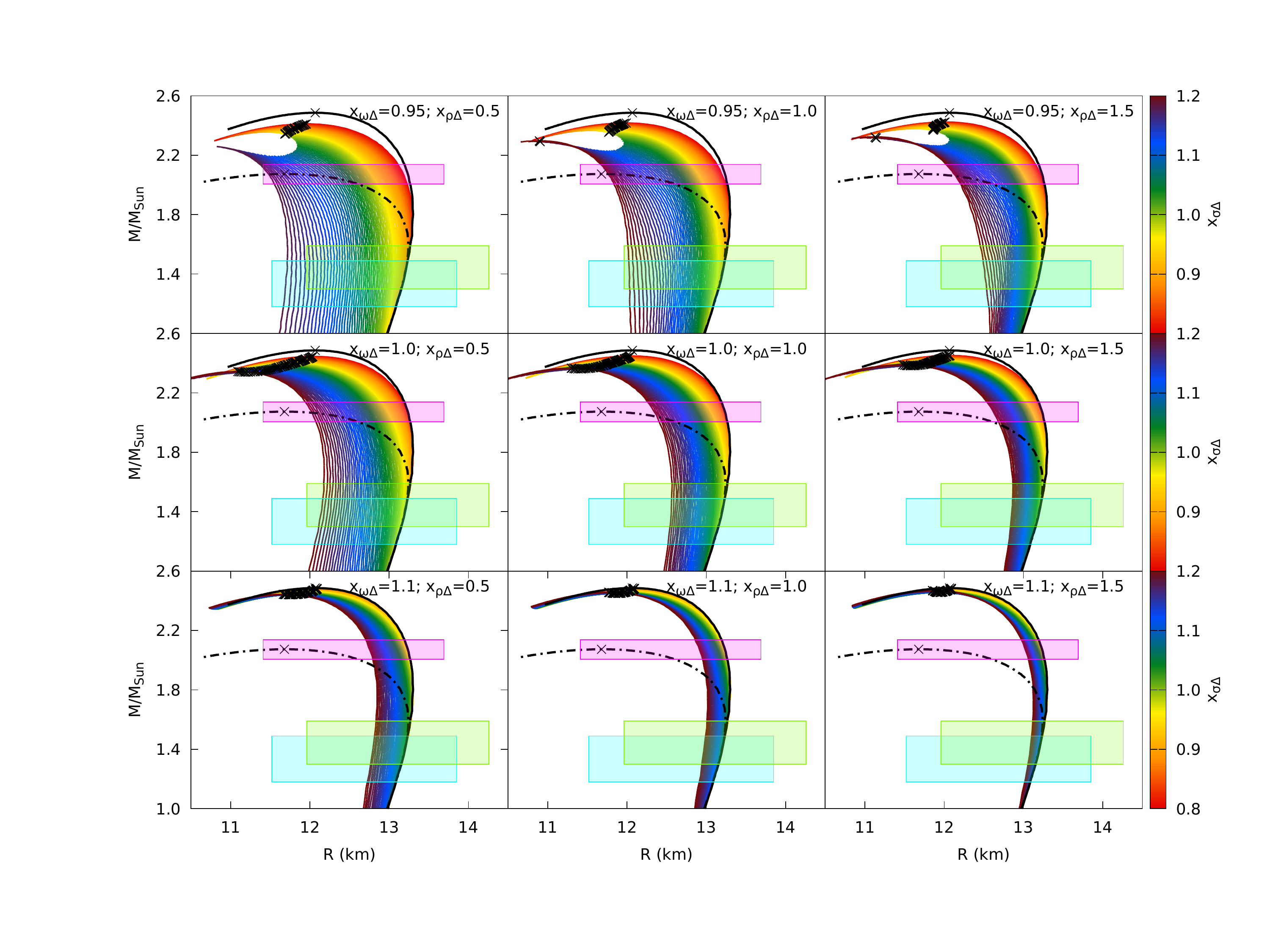}
\caption{Mass-radius diagrams for some choices of  $x_{\rho\Delta}$ and $x_{\omega\Delta}$, varying the $x_{\sigma\Delta}$ parameter for the N$\Delta$ matter composition {(color scale)}. The solid and dot-dashed black lines represent the N and NH compositions, respectively, and the black crosses indicate the maximum mass star if this configuration is reached. The {colored regions} represent NICER constraints (see text). }\label{fig:mr_nd}
\end{figure*}

\begin{figure*}[!t]
\centering
\includegraphics[width=\linewidth]{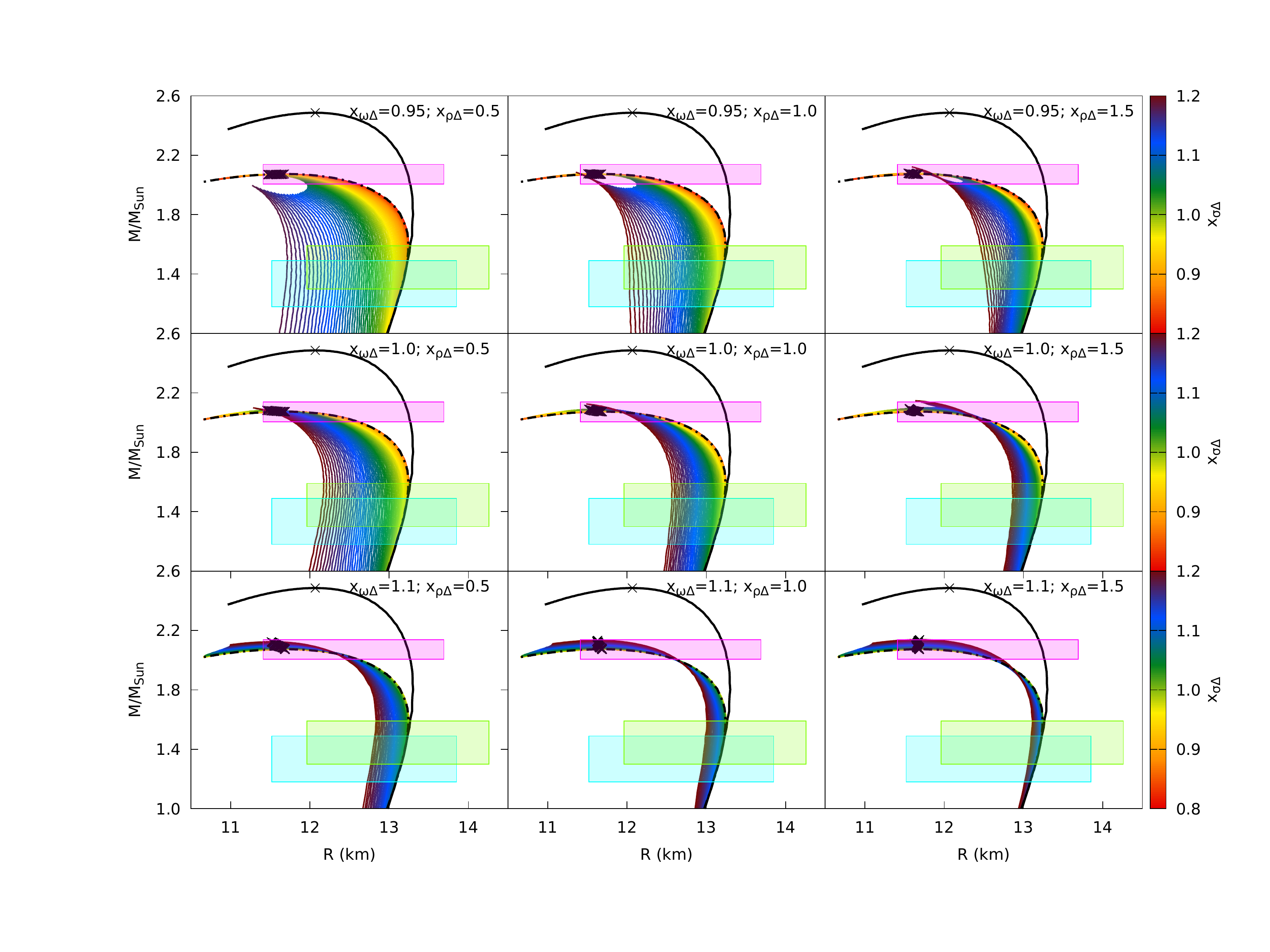}
\caption{Same as Fig. \ref{fig:mr_nd}, but for NH$\Delta$ matter composition.}\label{fig:mr_nhd}
\end{figure*}

The families of stars that result from the input of the obtained equations of state (EoS) in the Tolman-Oppenheimer-Volkoff (TOV) equations of relativistic hydrostatic equilibrium \cite{Tolman:1939jz,Oppenheimer:1939ne} are shown in Fig. \ref{fig:mr_nd} for hyperon-free matter and Fig. \ref{fig:mr_nhd} for $\Delta$-admixed matter including hyperons. In each figure we show results for three values of the  coupling-fractions $x_{\omega\Delta}$ (0.95, 1.0 and 1.1) and $x_{\rho\Delta}$ (0.5, 1.0 and 1.5). The colorbar indicates the  $x_{\sigma\Delta}$ value which we vary between 0.8 to 1.2. 
In the following figures, the full black line represents the results obtained with the pure nucleonic (N) EoS, and the black dash-dotted line has been calculated for a hyperonic (NH) EoS.
In these figures the crosses indicate the maximum mass configuration. 
The top panels in both figures, and middle panels of Fig. \ref{fig:mr_nhd}, show some EoS that do not reach the maximum mass star. In the presence of hyperons, this happens for  for $x_{\sigma\Delta}-x_{\omega\Delta}\gtrsim 0.1 $.
Formally, the maximum mass star is obtained when the TOV stability conditions of having a positive derivative of the star mass with respect to its central density ($\partial M/\partial\varepsilon_c\geq 0$) reaches a zero value. Black crosses indicate the maximum mass star for each EoS if this criteria is attained. As we will discuss later, some mass-radius curves do not reach the maximum configuration because the effective mass of the nucleon becomes zero at a too low density. 
This problem was identified in other works \cite{Kolomeitsev:2016ptu,Li:2018jvz,Raduta:2021xiz,Custodio:2022fbb}, but its consequences were not fully explored until now. 
In \cite{Kolomeitsev:2016ptu}, the authors have modified the model in order to avoid negative effective masses for the nucleon. 

The shaded {green and blue regions} in the figures represent observational constraints obtained from {two independent analysis of} NICER {data} 
of the pulsar {PSR J0030+0451}, that resulted in $M=1.34^{{}+0.15}_{{}-0.16}$ $\rm M_{\rm Sun}$ and $R=12.71^{{}+1.14}_{{}-1.19}$ km according to Ref. \cite{Riley:2019yda}, and in $M=1.44^{{}+0.15}_{{}-0.14}$ $\rm M_{\rm Sun}$ and $R=13.02^{{}+1.24}_{{}-1.06}$ km according to Ref. \cite{Miller:2019cac}, respectively. The magenta squared region represents the recent measurement of the pulsar {PSR J0740+6620} \cite{Cromartie2019,Fonseca:2021wxt} of $M=2.072^{{}+0.067}_{{}-0.066}$ $\rm M_{\rm Sun}$ and $R=12.39^{{}+1.30}_{{}-0.98}$ km, at a confidence interval of 68\% \cite{Riley:2021pdl}. 
{The uncertainties associated with the observations are not small enough to put strong constraints on the coupling parameters we are investigating. All models that reach the maximum mass configuration are compatible with the observational constraints for the several scenarios of matter composition considered, either with nucleons and $\Delta$s, or including hyperons as well. }

From the figures, we see that $x_{\sigma\Delta}$ competes with $x_{\omega\Delta}$ and $x_{\rho\Delta}$, with greater values of the first making the stellar radius decrease when compared with the $\Delta$-free matter composition. Larger values of $x_{\sigma\Delta}$ are associated with a larger attraction, and therefore a softer EoS at intermediate densities when the effect of the $\sigma$ meson dominates.  A similar effect occurs when smaller values of $x_{\rho\Delta}$ are taken: the smaller the $x_{\rho\Delta}$,  the smaller the radii obtained for a given pair $(x_{\sigma\Delta}$, $x_{\omega\Delta})$. This can be understood because a smaller  $x_{\rho\Delta}$ decreases the repulsion associated with the proton-neutron asymmetry. Another interesting effect is the fact that the simultaneous  presence of hyperons and $\Delta$s increases the maximum mass above the hyperonic matter maximum mass limit if $x_{\omega\Delta}\geq 1$. This is due to the fact that at high densities the effect of the vector meson dominates over the $\sigma$ meson and the $\Delta$ coupling to the $\omega$ meson is larger than the coupling of the nucleons or hyperons to the $\omega$-meson.
The role of the couplings in the maximum mass is quite complex, and will be better understood later in the discussion. 

\begin{figure*}[!p]
\centering
\includegraphics[width=\linewidth]{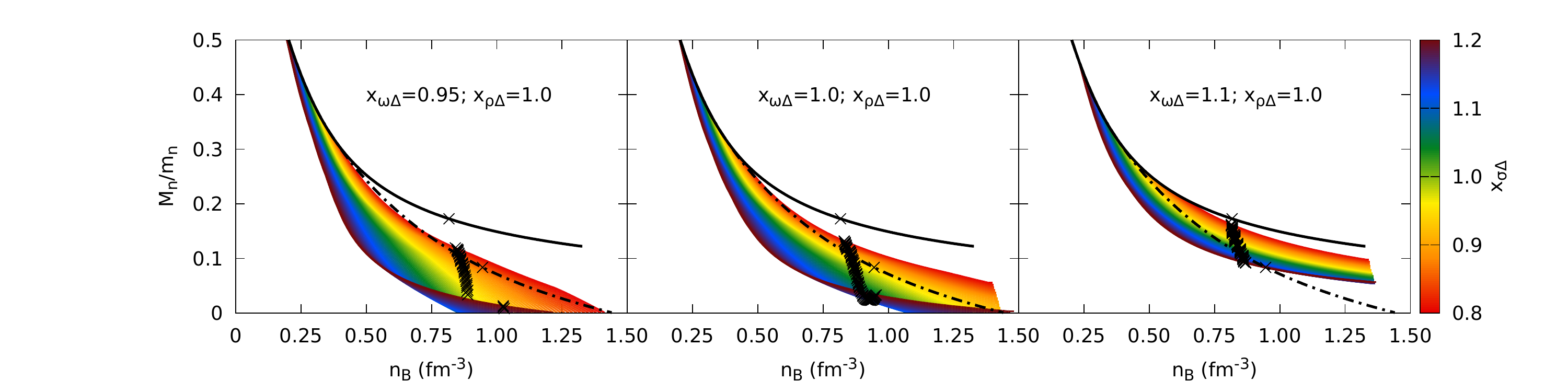}\\
\includegraphics[width=\linewidth]{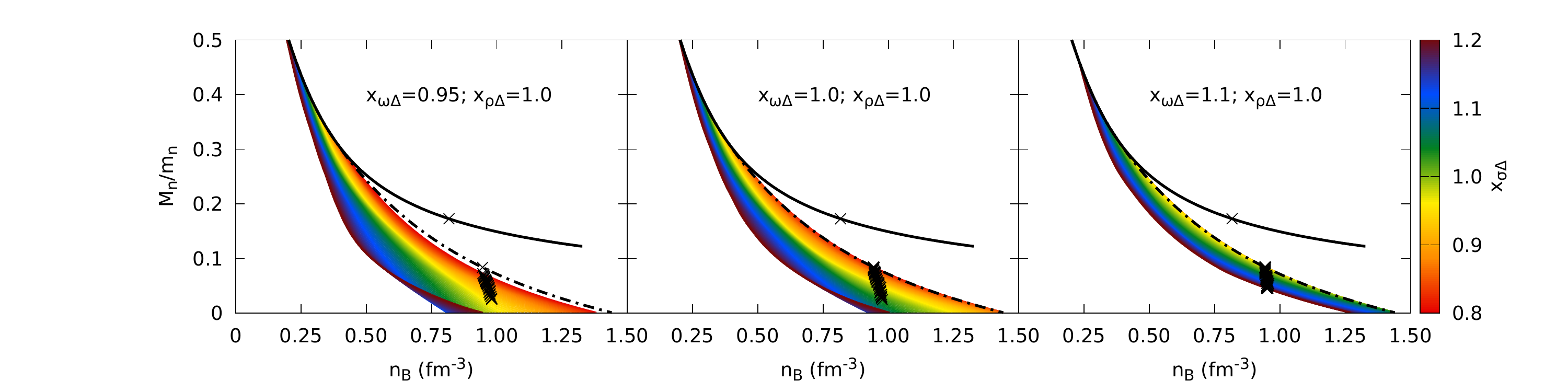} 
\caption{{Normalized} nucleon effective mass as a function of the density, taking $x_{\rho\Delta}=1.0$, for some choices of $x_{\omega\Delta}$ and varying the $x_{\sigma\Delta}$ parameter for the N$\Delta$ (top panels) and for NH$\Delta$  (bottom panels) matter compositions. The solid and dot-dashed black lines represent the N and NH compositions, respectively,  and black crosses indicate the central values of the maximum mass star if this configuration is reached.}\label{fig:mnuc_nd}

\centering
\includegraphics[width=\linewidth]{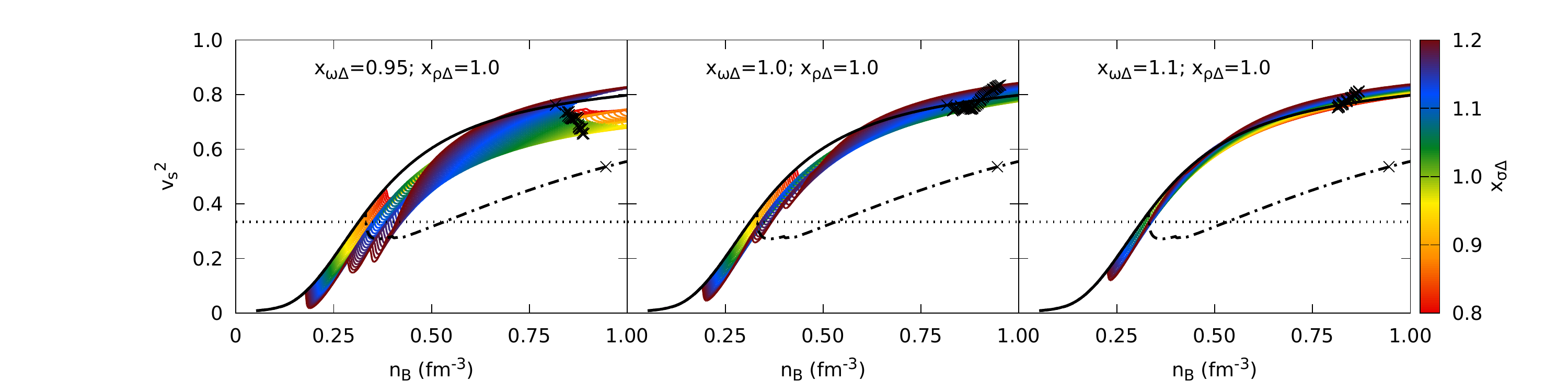}\\
\includegraphics[width=\linewidth]{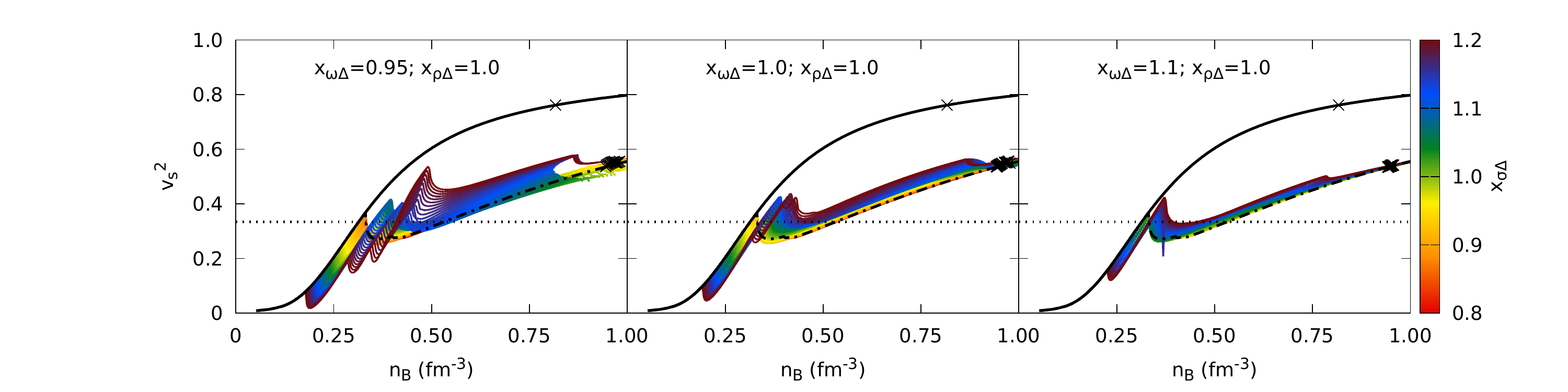}
\caption{Speed of sound squared as a function of the density, taking $x_{\rho\Delta}=1.0$, for some choices of $x_{\omega\Delta}$ and varying the $x_{\sigma\Delta}$ parameter for the N$\Delta$ (top panels) and NH$\Delta$ (bottom panels) matter composition. The solid and dot-dashed black lines represent the N and NH compositions, respectively, and the black crosses indicate the central values of the maximum mass star if this configuration is reached. The dotted line represents the conformal limit $v^2_s=1/3$.}\label{fig:vsound_nd}
\end{figure*}

In Fig. \ref{fig:mnuc_nd} we plot the nucleon effective mass,
\begin{equation}
    M_n=m_n-\Gamma_\sigma\sigma, \label{eq:meff}
\end{equation}
 as a function of the density. 
When we consider the nucleon-only neutron star matter composition, $M_n$ decreases asymptotically with $n_B$. When other baryon species are included in the matter composition (either hyperons, $\Delta$s, or both), we see a  much faster decrease of the nucleon effective mass.
This behavior is understood from the fact that each new particle present adds (through the scalar density dependence of the $\sigma$ field) to the negatively contributing term of Eq. \eqref{eq:meff}. 
The greater the multiplicity of baryons in the matter, the faster is the drop of $M_n$, as we can see from comparing  top and bottom panels of Fig. \ref{fig:mnuc_nd}  or even comparing the $\Delta$-admixed with the N or NH compositions inside each panel, noting that the higher values of $x_{\sigma\Delta}$ produce higher fractions of $\Delta$s.
For some configurations, the drop is so fast that the nucleon effective mass becomes too small and reaches zero before attaining the maximum densities expected to occur in the maximum mass configuration.
This behavior was already well-known for the hypernuclear star matter \cite{Schaffner:1995th}, but the inclusion of $\Delta$s makes it even more pronounced.
These EoS do not describe neutron stars properly, and therefore must be discarded from our analysis. We argue that these EoS would be valid if a phase transition to deconfined quark matter (or other exotic matter composition) could occur at a density below the one at which the nucleon effective mass becomes zero. This scenario will be explored in a future work.
For the models with  a non-vanishing effective nucleon mass, the EoS are computed while the thermodynamic stability condition $dP/d\varepsilon\geq0$ holds true. 
A liquid-gas type of phase transition is expected to occur when the thermodynamic stability is lost but, as the EoS can be computed to densities far beyond the ones present in stellar interiors (reaching at least $n_B=1.25$ fm$^{-3}$), and disregarding some unrealistic choices of very negative values of the relation ${x_{\sigma\Delta}-x_{\omega\Delta}}$, this behavior would not be prevalent in any physically reasonable scenario. 
 We will {disregard} the models that are not able to attain the maximum mass configuration when their EoS is applied to the TOV equations. {In a scenario that  allows for a hadron-quark deconfinement phase transition as in \cite{Drago:2014oja} but not considered in our study, they could still be acceptable.}
{We conclude that  the above results allow us to constrain   the $\Delta$ couplings due to some unphysical behavior such as the effective nucleon mass becoming zero at too low densities, or the EoS predicting a thermodynamic instability  near the saturation density that does not seem to be observed,  but  present known  astrophysical observations do not set any further constraint.} 
 
\begin{figure*}
\includegraphics[width=0.9\linewidth]{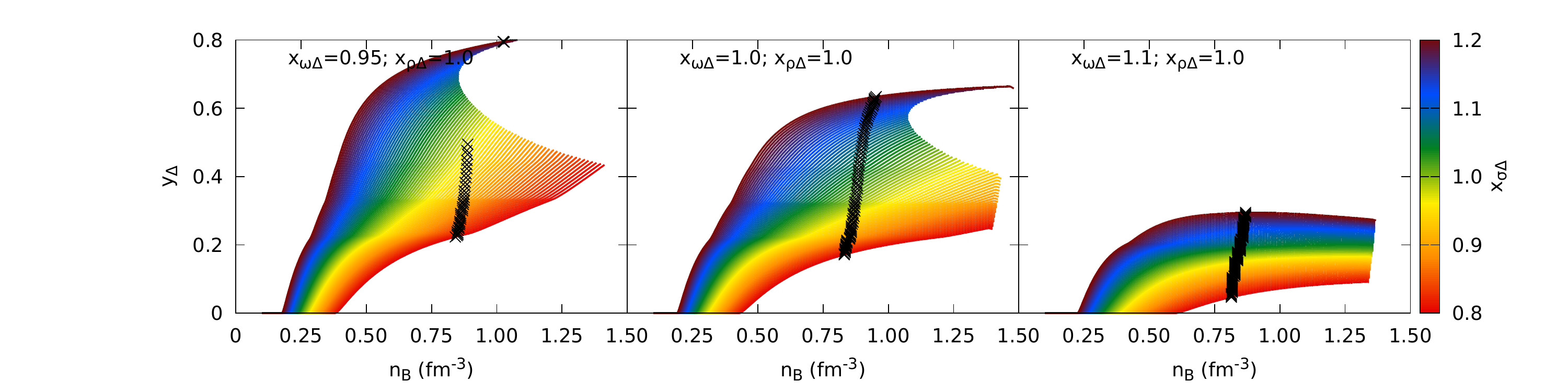}\\
\includegraphics[width=0.9\linewidth]{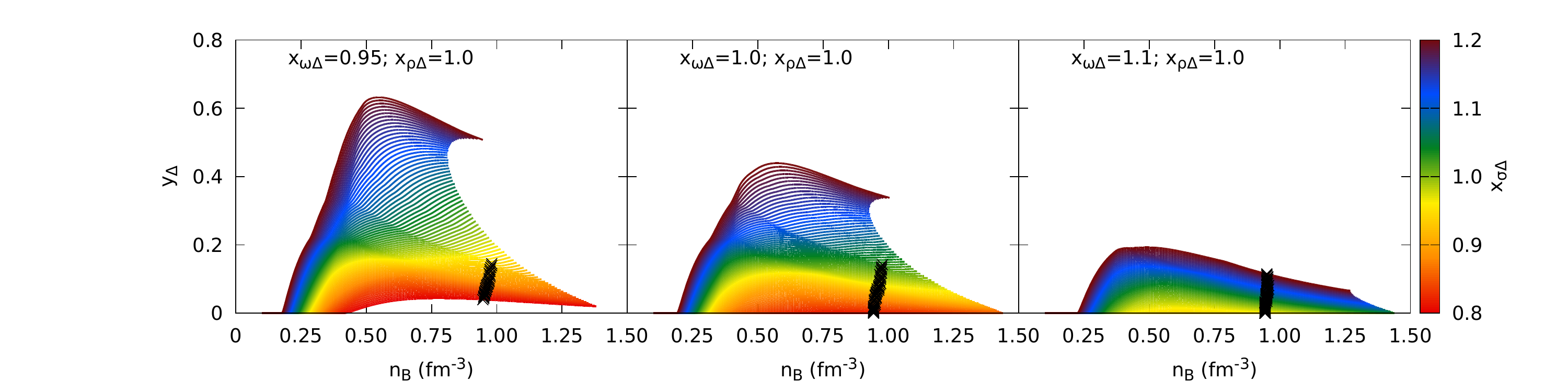}\\
\includegraphics[width=0.9\linewidth]{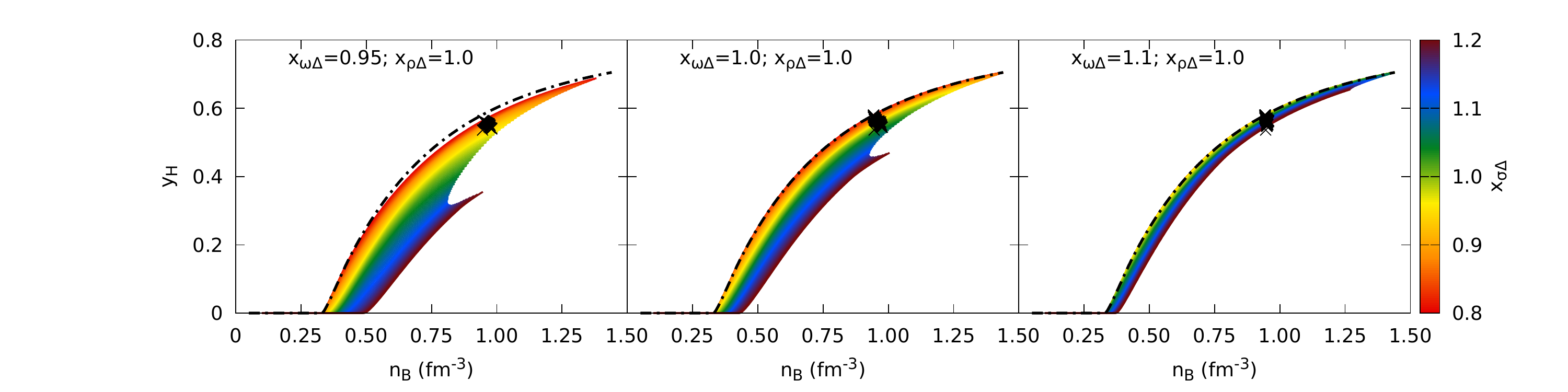}
\caption{Delta fraction  for the N$\Delta$ matter composition (top panels) and  for NH$\Delta$ matter composition (middle panels), and hyperon fraction for NH$\Delta$ matter (bottom panels) as a function of the density, taking $x_{\rho\Delta}=1.0$, for some choices of $x_{\omega\Delta}$ and varying the $x_{\sigma\Delta}$ parameter . Black crosses indicate the central values of the maximum mass star if this configuration is reached.}\label{fig:dfrac_nd}
\end{figure*}

 In Fig. \ref{fig:mnuc_nd}, the results are shown considering the whole computed EoS and, {as in the previous Figures,} black crosses indicate maximum mass star if this configuration is reached for the scenario in question.
 The maximum central density is around $n_B=0.85$ fm$^{-3}$ for the N$\Delta$ composition, and around $n_B=1.00$ fm$^{-3}$ for the NH$\Delta$ composition. 
 When $\Delta$s are favored to a point of suppressing all other species (higher values of $x_{\sigma\Delta}$ and/or lower values of $x_{\omega\Delta}$), the situation reverts back to the $N$ matter composition asymptotic behavior,  leading to the diminishing of the negatively contributing terms in Eq. \eqref{eq:meff},  but now  the $\Delta$ baryons are the most abundant particles.
 In this extreme limit, the EoS reaches the maximum mass star configuration once again, e.g., the indigo blue curve in Fig. \ref{fig:mnuc_nd} top left panel (this configuration is composed by a fraction of 80\% of $\Delta$s in the center of the star, see Fig. \ref{fig:dfrac_nd}).

The derivative of the pressure with respect to the energy density is the speed of sound, a quantity that provides information about shear viscosity, tidal deformability and gravitational waves signatures \cite{1988ApJ...333..880E,PhysRevC.102.055801,lopes2021neutron}. At zero temperature, its square is simply defined as
\begin{equation}
    v^2_s=\frac{\partial P}{\partial \varepsilon }~.
\end{equation}
It can be interpreted also as a measure of the EoS stiffness, with a higher speed generating a higher pressure at a given energy density and, therefore, sustaining a bigger star mass for a given radius. 
Results for the speed of sound are shown in Fig. \ref{fig:vsound_nd}, where one can notice the kinks due to the different particle onsets.
If only nucleonic matter is allowed in the N composition, quite high Fermi levels must be occupied. With the inclusion of new particles, the presence of more degrees of freedom distributes the Fermi pressure among the different particles and  softens the EoS. It holds true in the intermediate densities (for $n_B<0.50$ fm$^{-3}$) for the N$\Delta$ composition, and always after the hyperon onset in the NH and NH$\Delta$ compositions.
The behavior of hyperonic neutron-star matter, however, is affected in a more complicated way by the inclusion of $\Delta$ baryons. The NH$\Delta$ composition is softer than the NH case at lower densities, but this situation is reversed at the middle regions. This is due to the strong coupling of the $\Delta$s to the $\omega$ meson. For the same reason at high densities, N$\Delta$ matter has a larger speed of sound than N matter. This difference is then reduced in the higher densities once again. 

{Perturbative QCD results for very high densities (more than 40 times the nuclear saturation density) predict an upper limit of $v^2_s=1/3$ \cite{Bedaque:2014sqa,Annala:2019puf}. In such high densities, far beyond the ones
reached in the neutron star interiors, the baryonic matter is expected to be deconfined in quark matter. 
However, several authors have discussed that the two solar mass constraint requires a speed of sound well above the conformal limit, indicating that matter inside NS is a strongly interacting system \cite{Bedaque:2014sqa,Alford:2013aca,Moustakidis:2016sab,Tews:2018kmu,Reed:2019ezm}. 
Nevertheless, within the description undertaken in \cite{Annala:2019puf}, it was shown that the size of the quark core in hybrid stars is related to the speed of the sound of the quark matter, and very massive quark matter cores are expected in the NS interiors if the conformal limit is not strongly violated. 
As shown in Fig. \ref{fig:vsound_nd}, the onset of hyperons and $\Delta$s breaks the monotonic behaviour of $v_s^2$, reducing the speed of sound, but the conformal limit is always violated due to the fact that  we are describing hadronic (and not deconfined quark) matter.
The speed of sound behavior, the sudden decrease,   is  similar to the one found in other works when new degrees of freedom set in, such as the onset of  hyperons  in \cite{lopes2021nature} or of  $s$ quarks in \cite{Ferreira:2021osk}.}

The relative populations of each kind of baryons are shown in Fig. \ref{fig:dfrac_nd}, where we have defined the particle fractions as $y_i=\sum_b n_b/n_B$, with $i=\{H,\Delta\}$, meaning that the summation runs only over the hyperons or $\Delta$s, respectively. 
Very large $\Delta$ fractions are expected for the larger values of $x_{\sigma\Delta}$, the effect being quite drastic if $x_{\omega\Delta}<1$. In this case, many EoS do no attain the maximum-mass star, and are considered invalid. In the presence of hyperons, the condition of attaining the maximum mass configuration is stronger, because the nucleon effective mass goes to zero too soon.  Taking $x_{\omega\Delta}>1$, these difficulties cease to occur.
The hyperon fractions are also shown in Fig. \ref{fig:dfrac_nd}  bottom panels. As expected, larger $x_{\sigma\Delta}$ couplings, which favor the appearance of $\Delta$s, will disfavor the appearance of hyperons. This completes the conclusion drawn from Fig. \ref{fig:dfrac_nd} middle panels where it is seen that for stars  with both $\Delta$s and hyperons, large $\Delta$ contents  do not reach the maximum-mass configuration.  We also conclude that for models that are able to attain the maximum-mass configuration identified by the cross, the hyperon fraction at the center of the star is of the order of 50\% and the $\Delta$ fraction is below 20\%. In the presence of hyperons, the maximum $\Delta$ fraction is attained for densities between 2$\rho_0$ and 3$\rho_0$ and takes values below 30\%. Although the $\Delta$ baryons set in first, they are replaced by hyperons at high densities because the coupling of the $\Delta$ baryons to the $\omega$ meson is stronger.
Looking for, e.g., the upper-mid panel of Fig. \ref{fig:mr_nd}, we identify an isolated configuration where the EoS reaches the maximum mass with a very large $x_{\sigma\Delta}$.
From the  top left panel of Fig. \ref{fig:dfrac_nd}, it is possible to see that this configuration is composed of around $80\%$ of $\Delta$ baryons, considering all isospin projections together.
It explains why the nucleon effective mass reverts to the asymptotic behavior in order to allow the description at higher densities (see the left panel of Fig. \ref{fig:mnuc_nd}).
These results suggest that compact stars might exist in some hyperon-free $\Delta$-dominated composition, that we label \textit{deltic stars}.

\begin{figure}[!t]
\centering
\includegraphics[width=\linewidth]{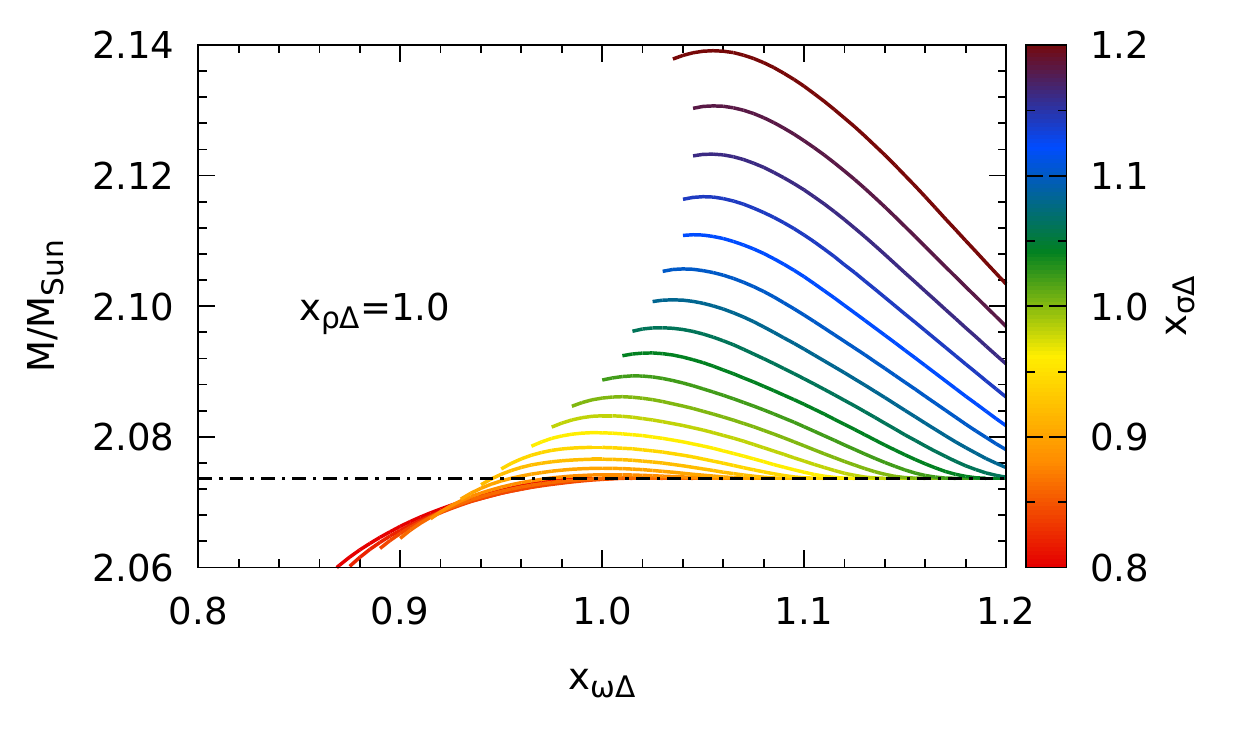}
\caption{Maximum stellar mass as a function of the $x_{\omega \Delta}$ coupling, taking $x_{\rho\Delta}=1.0$, varying the $x_{\sigma\Delta}$ parameter for the NH$\Delta$ matter composition. The dot-dashed black horizontal line represents the maximum mass for the NH composition ($M=2.07$ M$_{\rm Sun}$), and the curves are plotted only for values where the maximum mass star configuration is reached. }\label{fig:mmax}
\end{figure}

The bottom row of Fig. \ref{fig:mr_nd} allows us to see a rather unexpected behavior. For these choices of $x_{\omega\Delta}$ and $x_{\rho\Delta}$, the maximum masses increase with $x_{\sigma\Delta}$, i.e., with a greater fraction of $\Delta$s (see Fig. \ref{fig:dfrac_nd}). It may be considered counter intuitive since, taking as an example  the hyperon puzzle \cite{Chamel:2013efa}, the inclusion of more particles involves more degrees of freedom, and  lowers the Fermi levels. Following this reasoning, it is expected that the admixture of $\Delta$s in hypernuclear matter would make the EoS softer, but it is not always the case.
In Walecka-type relativistic models (a category in which we include the DDME2 and other density-dependent parameterizations in), the attractive $\sigma$ field grows rapidly until about 3 times the saturation density, but then shows a softer dependence on $n_B$ at higher densities. 
On the other hand, the repulsive $\omega$ field grows indefinitely in a {linear} fashion and, then, becomes dominant in the denser regions. 
There are more $\Delta$s in the matter composition for larger $\sigma$-delta couplings, and, since the $\omega$-delta coupling is always taken to be much greater than the $\omega$-hyperon coupling (that is not greater than $\sim 2/3 g_{\omega}$), configurations where $\Delta$s are more abundant will have a stronger repulsion than scenarios that only consider the NH composition, resulting in a stiffer EoS and a higher maximum mass.

{The effect of the delta-meson couplings on the maximum stellar mass is illustrated in Fig. \ref{fig:mmax}.
We note that, for a fixed $x_{\omega \Delta}$, increasing the parameter $x_{\sigma \Delta}$ will always produce a more massive star.
When the parameter $x_{\sigma \Delta}$ is fixed, the maximum mass will reduce slightly for greater $x_{\omega \Delta}$.
The main factor in play here is the balance between the relative fractions of hyperons and $\Delta$s: a larger $x_{\sigma \Delta}$ favors larger $\Delta$ fractions. The $\Delta$s couple more strongly to the  $\omega$ fields.  Even though, stronger $\sigma-$meson couplings are involved, the $\omega$ field dominance at large densities results in a  stiffening of the EoS, and, therefore, larger masses. This balance is stronger for $1.0<x_{\omega \Delta}<1.2$.
 In \cite{Li:2018jvz}, a similar conclusion was drawn, although the maximum mass was obtained for $1.1<x_{\omega \Delta}<1.2$, implying  smaller values. 
 This difference is probably occurring because a different hyperon interaction was considered. Notice, however, that we do not consider $x_{\omega\Delta}>1.2$ and that with our parametrization we do not get maximum mass configurations for  $x_{\sigma\Delta}-x_{\omega\Delta}\gtrsim 0.1 $.
}

\begin{figure}[!t]
\centering
\includegraphics[width=\linewidth]{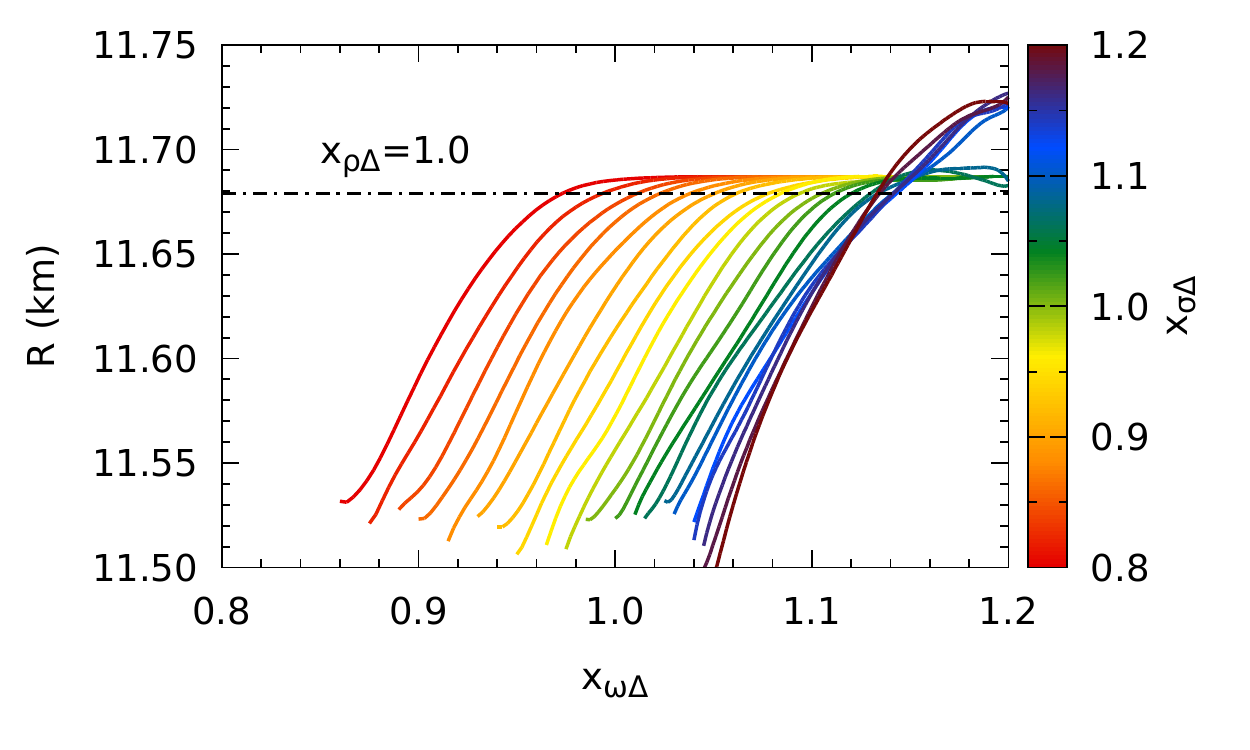}\\
\includegraphics[width=\linewidth]{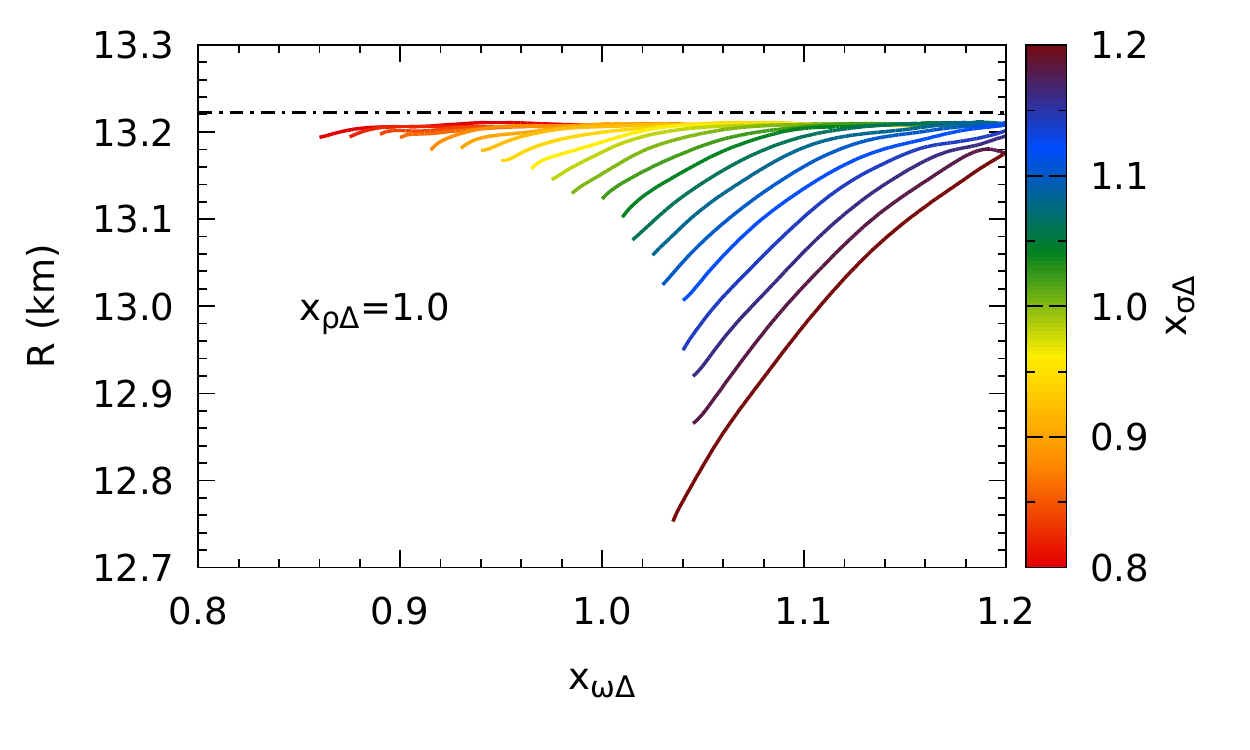}
\caption{Radii of the maximum mass (top) and canonical (bottom) stars as a function of the $x_{\omega \Delta}$ coupling, taking $x_{\rho\Delta}=1.0$, varying the $x_{\sigma\Delta}$ parameter for the NH$\Delta$ matter composition. The dot-dashed black horizontal lines represent the radii of the NH matter stars  ($R=11.67$ km and $R=13.22$ km, respectively), and the curves are plotted only for values where the maximum mass star configuration is reached. }\label{fig:radii}
\end{figure}

In Fig. \ref{fig:radii}, we perform a similar study for the radius of the maximum mass star (top panel) and radius of the 1.4 M$_{\rm Sun}$ star (bottom panel).   For $x_{\rho\Delta}=1.0$, the presence of $\Delta$s may reduce the maximum mass radius in 100-150 m,  $11.5$ km being the minimum, and the $1.4$ M$_{\rm Sun}$ star radius in 20-500 m, with a minimum of $12.7$ km. 
Only models that attain maximum mass configurations are represented in Fig. \ref{fig:mmax} and  \ref{fig:radii}.
In \cite{Li:2018qaw} smaller radii are obtained, and the presence of $\Delta$s may give rise to a reduction of the radius of the canonical star  of up to  $\approx 2$ km. However, it is not clear if the authors exclude models that do not attain the maximum mass. In \cite{Ribes:2019kno}, the authors have obtained,  with FSU2H, effects of the order of the ones discussed in the present work with DDME2.
From Fig. \ref{fig:mr_nd} and  \ref{fig:radii}, it is seen that the presence of $\Delta$s (induced by larger values of $x_{\sigma\Delta}$) cause a significant decrease in the radius of the stars with intermediate masses. 
This is explained by the fact that the appearance of the $\Delta$s softens the EoS in the intermediate density region, as clearly seen in the top panels of Fig. \ref{fig:vsound_nd}.
{Stars  with core densities in this range are further compressed when $\Delta$s set in  and, consequently, their radii reduce \cite{Ribes:2019kno}.}

\begin{figure*}[!p]
\centering
\includegraphics[width=\linewidth]{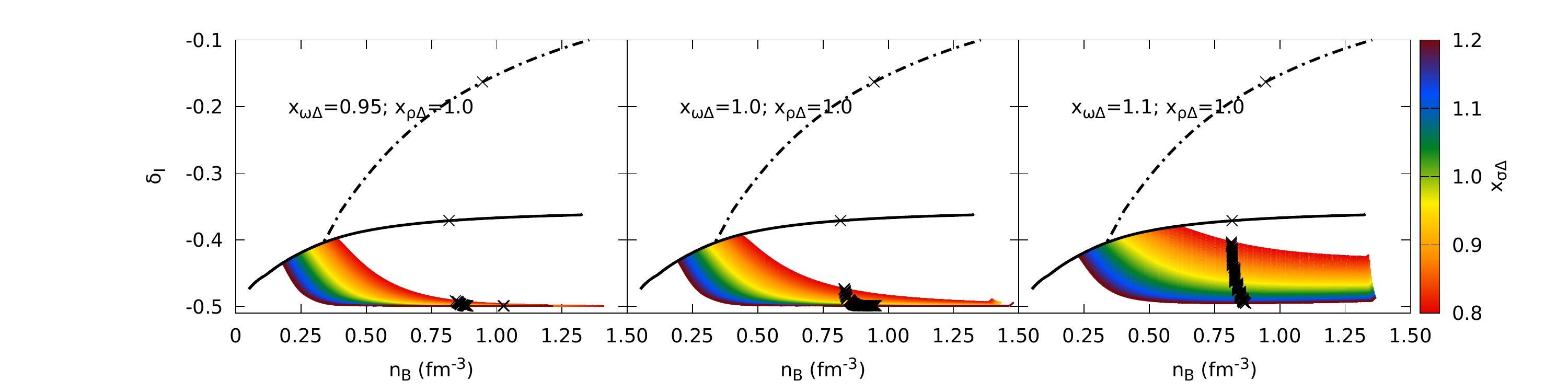}\\
\includegraphics[width=\linewidth]{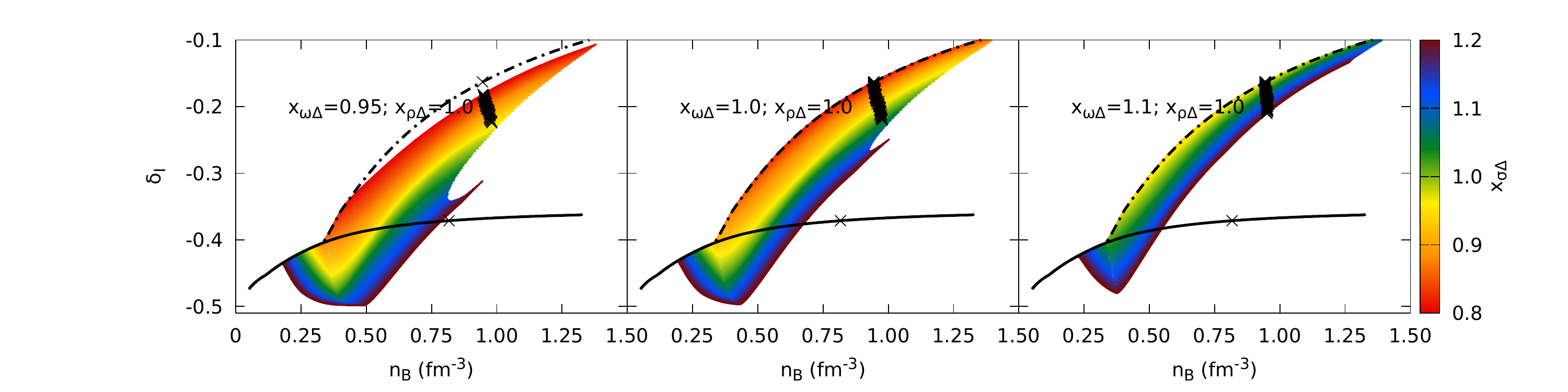}
\caption{Isospin asymmetry coefficient Eq. \eqref{eq:rhoiz} as a function of the density, taking $x_{\rho\Delta}=1.0$, for some choices of $x_{\omega\Delta}$ and varying the $x_{\sigma\Delta}$ parameter for the N$\Delta$ matter composition (top panels) and for NH$\Delta$ matter composition (bottom panels). The solid and dot-dashed black lines represent the N and NH compositions, respectively.  }\label{fig:rhoiz_nd}

\centering
\includegraphics[width=1.0\linewidth]{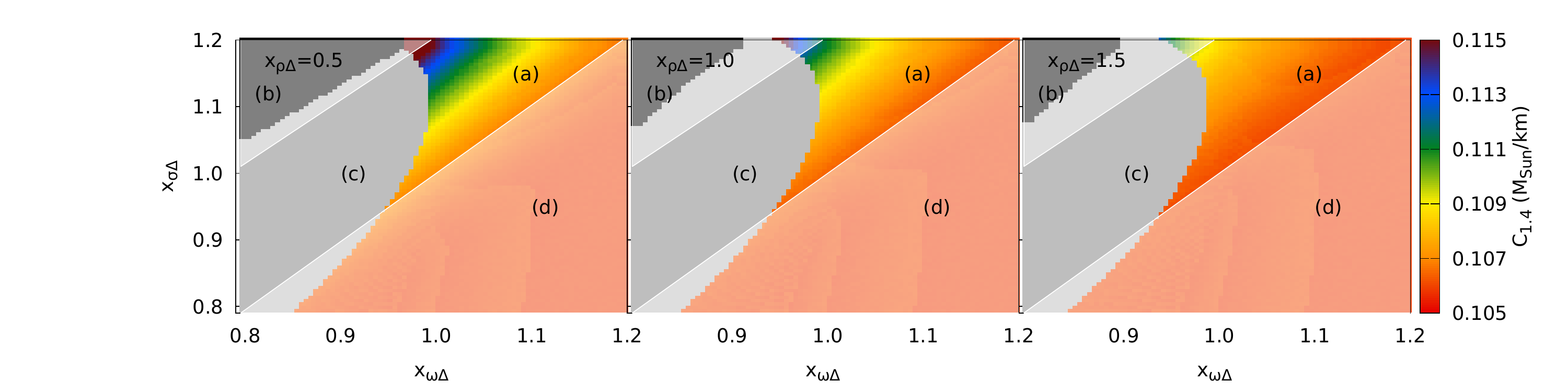}\\
\includegraphics[width=1.0\linewidth]{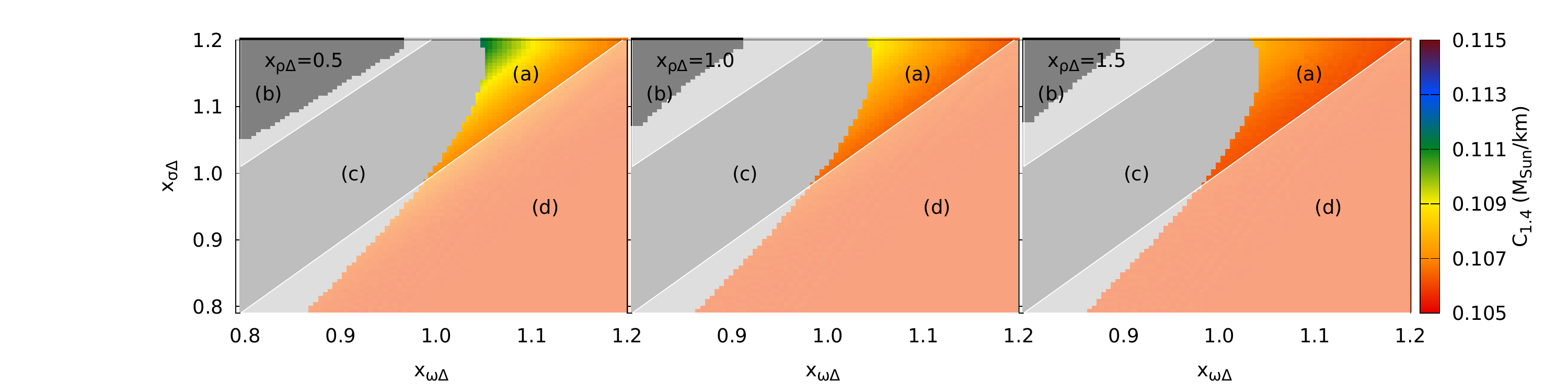}
\caption{Constraints on meson-delta couplings, for the N$\Delta$ matter composition (top panels) and  for the  NH$\Delta$ matter composition (bottom panels). The color gradient (a) indicates the compactness of the canonical star.
The black region (b) represents values for which $\frac{dP}{d\varepsilon}<0$ before $n_B=0.2$ fm$^{-3}$.
The gray region (c) represents values where the maximum mass star configuration is not reached when inputting the EoS in the TOV equations.
The white-shaded region (d) indicates the combinations of
parameters that do not fulﬁll the constraint given by Eq. \eqref{eq:constr} \cite{WEHRBERGER1989797}. 
}\label{fig:crit_nd}
\end{figure*}

The stiffening of the EoS due to the $\Delta$ admixture, was also noticed in Ref. \cite{Dexheimer:2021sxs}, where it was suggested that it occurs due to the isospin asymmetry.
We  define the coefficient 
\begin{equation}
    \delta _{I}=\frac{\sum_b I_{3\,b}n_b}{\sum_b n_b}~,\label{eq:rhoiz}
\end{equation}
that represents the average 3$^{\rm rd}$ isospin component of a given matter composition, weighted by each particle relative density, as shown in Fig. \ref{fig:rhoiz_nd}.
The density at which the curves with and without $\Delta$s split marks the appearance of the $\Delta^-$ baryon, which turns the isospin asymmetry more negative and, consequently, makes the EoS stiffer. It happens earlier for larger $x_{\sigma\Delta}$ couplings, as this is the determinant parameter to favor the onset of the $\Delta$s.
For the N$\Delta$ composition (top panels of Fig. \ref{fig:rhoiz_nd}), the isospin asymmetry coefficient tends to more negative values as the density increase, because the matter turns to be dominated by the $\Delta^-$. This tendency is stronger for smaller $x_{\omega\Delta}$ couplings, as a strong $\omega$ coupling does not favor $\Delta$ populations at higher densities. 
However, when the NH$\Delta$ composition is considered (bottom panels of Fig. \ref{fig:rhoiz_nd}), the isospin asymmetry coefficient becomes less negative once the hyperon threshold is reached and follows the NH composition behavior after that, becoming less negative as the matter is more dominated by the hyperons.
Nevertheless, the configurations with relatively more $\Delta$s present (i.e., bigger $x_{\sigma\Delta}$, drawn in indigo blue in the plots) show more negative values of $\delta _{I}$.

In Fig. \ref{fig:crit_nd}, we summarize the constraints on the values of the couplings that ensure the existence of neutron stars compatible with the stability criteria and with the observational results.  For three values of the coupling $x_{\rho\Delta}$ (0.5, 1 and 1.5), the compactness of a 1.4 M$_{\rm Sun}$ star is plotted versus the $x_{\sigma\Delta}$ and $x_{\omega\Delta}$. 
The color gradient indicates the compactness, defined as $C_M={M}/{R}$, of the canonical ($M=1.4$ M$_{\rm Sun}$) star. The compactness of the isolated neutron star {RX J0720.4-3125} is inferred to be $C=0.105\pm 0.002$ {$M_{{\rm Sun}}/$km} 
\cite{Hambaryan:2017wvm}, which gives us an additional parameter for analysis, specially focused on the less massive star radii. 
From the figures, we  see that the effect of $x_{\rho\Delta}$ is making the canonical star less compact as the parameter increases, improving the agreement with this constraint.
The black region on the upper-left corner represents values for which ${dP}/{d\varepsilon}<0$ before $n_B=0.2$ fm$^{-3}$, meaning that the thermodynamic stability condition is not satisfied at these low densities.
The gray region represents values where the maximum mass star configuration is not reached because the effective nucleon mass goes to zero. Note that all configurations approved by these two criteria  fulfill the observational constraints shown in Fig. \ref{fig:mr_nd}. 
The white-shaded triangular regions indicate the combinations of parameters that do not fulﬁll the constraint given by Eq. \eqref{eq:constr} \cite{WEHRBERGER1989797}.
The remaining points, {identified with (a) and indicated by the color gradient},  correspond to delta-meson couplings that satisfy all constraints.
{Comparing with the coupling domain obtained in \cite{Ribes:2019kno}, {in this work} a larger domain was obtained, indicating that solutions with $ x_{\omega\Delta>1.0}$ and $ x_{\sigma\Delta}>1.0$  are possible. The difference is essentially connected with the model: DDME2 allows for a larger parameter domain for which the effective mass does not go to zero before the maximum mass configuration is attained. For a large enough $ x_{\rho\Delta}$, the constraint  $C=0.105\pm 0.002$ $M_{{\rm Sun}}/$km is satisfied for $ x_{\sigma\Delta}$ and $ x_{\omega\Delta}$ larger than one. A smaller value of $ x_{\rho\Delta}$, e.g.  0.5 in the left panel, is more constraining with respect to the combination $ x_{\sigma\Delta}$-$ x_{\omega\Delta}$ and does not allow for large values of $ x_{\sigma\Delta}$.}

\begin{figure*}[!t]
\centering
\includegraphics[width=1.\linewidth]{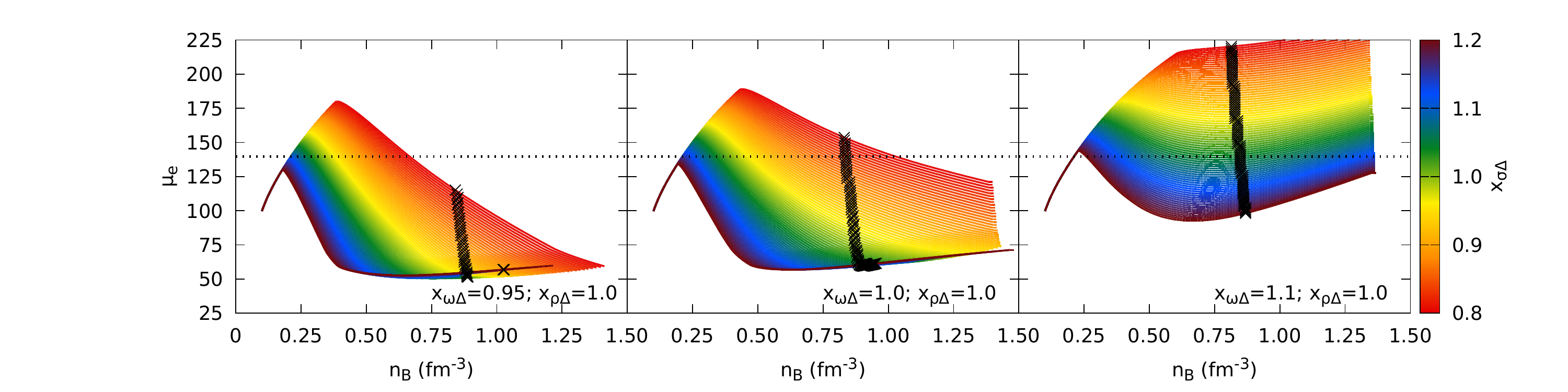}\\
\includegraphics[width=1.\linewidth]{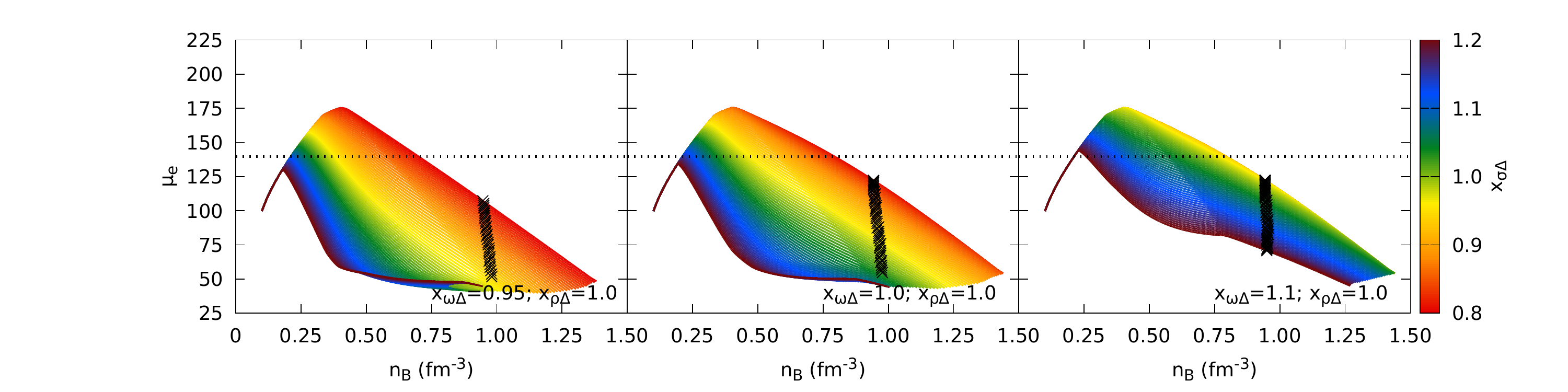}
\caption{{Electron} chemical potentials as a function of the density, taking $x_{\rho\Delta}=1.0$, for some choices of $x_{\omega\Delta}$ and varying the $x_{\sigma\Delta}$ parameter  for  N$\Delta$ (top panels) or NH$\Delta$ (bottom panels) matter composition. The dashed line represents the vacuum $\pi$ mass. }
\label{chempot}
\end{figure*}

{In the present work, we have considered that the $\Delta$ baryons are stable in stellar matter, as considered in many other works \cite{Drago:2014oja,Li:2018jvz,Raduta:2021xiz}. This may be justified because the possible final states for the decay of the $\Delta$ to occur are blocked. 
Generally, in vacuum the deltas $\Delta=\{\Delta^{++},\Delta^{+},\Delta^{0},\Delta^{-}\}$ quickly decay via the strong force into a nucleon $N=\{p,n\}$ and a pion of appropriate charge, 
$$\Delta\longrightarrow N+\pi.$$ 
The $\Delta^-$ will decay if $\mu_{\Delta^-}-\mu_n= \mu_{\pi^-}$, giving possibly rise  to a $\pi^-$-condensate,
{and} in $\beta$-equilibrium $\mu_{\pi^-}=\mu_e$.
For reference, we show in Fig. \ref{chempot} the electron chemical potential (i.e., the difference between the $\Delta$ and the nucleon (effective) chemical potentials), and indicate the vacuum pion mass with a dotted line, $m_\pi=139.5$ MeV. 
In several scenarios the electron chemical potential is larger than the pion vacuum mass. {This would indicate that indeed the pion condensate would be favorable.}
However, {in Refs.~\cite{Glendenning:1984jr,Ohnishi:2008ng}, the authors showed} that the repulsive  s-wave pion-nucleon interaction does not favor pion condensation {because, in the medium, the pion energy is above its vacuum mass.} 
}

\section{Conclusions}

By applying a  relativistic mean-field formalism, we have analysed the properties of NS  matter with an admixture of nuclei and $\Delta$ baryons, and an admixture of nuclei, hyperons and $\Delta$ baryons.  The meson-hyperon couplings were chosen by imposing hypernuclear properties \cite{Fortin:2017cvt,Fortin:2020qin}. For the $\Delta$-meson couplings, we have considered the constraint obtained in \cite{WEHRBERGER1989797} from electron-nucleon measurements.

It was shown that under some conditions, in particular, in the presence of a large admixture of different particles the  nucleon effective mass goes to zero before the maximum mass configuration is reached. These configurations were considered unphysical. It was shown that using astrophysical observations to constraint the couplings, the onset of the $\Delta^-$ baryon is favored over the hyperons, in particular the $\Lambda$ hyperon.  Some  NS configurations were determined with $\approx$80\% $\Delta^-$ baryons in the core center, the so called {\it deltic stars}.  It was shown that at large densities the presence of $\Delta$s would generate a quite stiff  EoS due to the $\omega$ meson dominance. As a consequence, NS  with a nucleon-hyperon-delta admixure attains larger maximum masses and a larger speed of sound in the core center.  For the same reason, in the presence of hyperons the $\Delta$ distribution is maximum at intermediate densities, below 30\%, and reduces towards the NS star center to values below 20\%.

Taking as reference the isolated neutron star {RX J0720.4-3125}, for which the compactness $C_M=0.105\pm 0.002$ $M_{{\rm Sun}}/$km \cite{compactness} has been measured, it was shown that values of $x_{\sigma\Delta}$, $x_{\omega\Delta}$, and $x_{\rho\Delta}\gtrsim 1.0$ are favored together with $x_{\sigma\Delta}>x_{\omega\Delta}$.  A small $x_{\rho\Delta}$ favors smaller radii in neutron stars with intermediate masses.

The calculation of the electron chemical potential has allowed us to discuss the possibility of the occurrence  of a pion condensate. However, the values that we have obtained were never much larger than the pion vacuum mass, which may indicate that its condensation may not be favoured. Also, and according to Refs.~\cite{chen2010,Ohnishi:2008ng}, this possibility was disfavored because the pion s-wave interacts repulsively with nucleonic matter.

Finally,  all stars {obtained in this work} with a mass above two solar masses satisfy the NICER constraints for pulsars {PSR J0740+6620} and {PSR J0030+0451}. The maximum NS masses with an admixture of hyperons and $\Delta$s was obtained for $x_{\sigma\Delta}=1.2$ and $x_{\omega\Delta}\sim 1.05$, corresponding to a maximum fraction of $\Delta$s: to obtain the maximum fraction,  a large $x_{\sigma\Delta}$  is needed but not  a too large $x_{\omega\Delta}$. The presence of $\Delta$s may originate, for the canonical mass, a  reduction {in the radius} of up to 500m,  and of 200m for the maximum-mass configuration, {and considering} $x_{\rho\Delta}=1$. Taking a smaller value, e.g. $x_{\rho\Delta}=0.5$, the radius of the canonical star may be reduced up to $\approx 1$km.  In the present work, we have obtained a larger coupling domain than the one determined in \cite{Ribes:2019kno}, but certainly much smaller than the one  discussed in \cite{Li:2018qaw}.

\section*{Acknowledgements}
This work is a part of the project INCT-FNA Proc. No. 464898/2014-5, {and also partly supported by the FCT (Portugal) Projects No. UIDB/FIS/04564/2020, UIDP/FIS/04564/2020, and by PHAROS COST Action CA16214}. K.D.M. acknowledges a doctorate scholarship from Conselho Nacional de Desenvolvimento Científico e Tecnológico (CNPq/Brazil). D.P.M. was partially supported by Conselho Nacional de Desenvolvimento Científico e Tecnológico (CNPq/Brazil) under grant 303490-2021-7. H.P. acknowledges the grant CEECIND/03092/2017 (FCT, Portugal).

\bibliography{references}

\end{document}